\documentclass{iopjournal}

\usepackage{amsmath,amssymb,amsfonts,dsfont,xspace,graphicx,relsize,bm,mathtools,xcolor,amsthm,soul}
\usepackage{graphicx}
\usepackage{dcolumn}
\usepackage{xcolor}
\usepackage{bbm}
\usepackage{physics}
\usepackage{ulem}
\usepackage{hyperref}
\hypersetup{
	colorlinks = true,
	urlcolor   = blue,
	linkcolor  = blue,
	citecolor  = magenta
}
\usepackage{algorithm}
\usepackage{algpseudocode}
\usepackage{float}
\usepackage[shortlabels]{enumitem}
\usepackage{subfigure}
\usepackage{orcidlink}
\usepackage{subcaption}
\usepackage{siunitx}
\usepackage{rotating} 
\usepackage{tabularx} 
\usepackage{array}
\usepackage{adjustbox}
\usepackage{url}

\begin{document}

\articletype{Paper} 

\title{Engineering of Chirped Apodized Sources for Quantum Spectroscopy}

\author{Linda Sansoni$^{1,2}$\orcid{0000-0002-4445-1036}, Max Widarsson$^{3}$\orcid{0000-0002-2397-0494}, Patrick Mutter$^{3}$\orcid{0000-0002-2526-4921} and Andrea Chiuri$^{1,2,*}$\orcid{0000-0001-9733-0740}}

\affil{$^1$ENEA - Nuclear Department, Via E. Fermi 45, 00100 Frascati, Italy}

\affil{$^2$INFN Sezione Roma Tre, Via della Vasca Navale, 84, 00146 Rome, Italy}

\affil{$^2$SLF Quantum, part of SLF Svenska Laserfabriken AB, Ruddammsvägen 49, 114 19 Stockholm, Sweden}

\affil{$^*$Author to whom any correspondence should be addressed.}

\email{andrea.chiuri@enea.it}


\begin{abstract}
Quantum sources of light are of paramount importance for the development of photonic quantum technologies. The growing demand for bright, efficient, and spectrally engineered photonic sources for applications in quantum sensing, spectroscopy, and imaging has motivated the development of tailored nonlinear crystals. Here, we report on the development and characterization of chirped, apodized KTP crystals specifically engineered for quantum spectroscopy. Our source exhibits an unprecedented bandwidth in a highly non-degenerate configuration, demonstrating its potential to support quantum spectroscopy protocols over broad spectral ranges while enabling significantly shorter measurement times.
\end{abstract}

\section{Introduction}

Quantum-light sources have been increasingly investigated for spectroscopy, imaging, sensing \cite{Basset19,Fuenzalida24,Defienne2024}, and remote measurements \cite{Pitsch24SPIE,Neves26}, exploiting spectral \cite{Hashimoto25,Chiuri2025,Placke26}, temporal \cite{Denis2017}, spatial \cite{padgett17ptrsa}, and polarization \cite{Magnitskiy22} correlations between photons. Within this broad panorama of possible applications, this manuscript will be focused on the Quantum Spectroscopy, the counterpart of the Quantum Imaging in the frequency domain, that is a crucial topic within the Quantum Technologies. Here the sources of quantum light represent pillar components to increase their performances and to fit the requirements arising from the market and the desired applications. Indeed, the Quantum Spectroscopy relies on specific technical specifications requisites in terms of bandwidth, brightness of the source and the spectral shape of the emitted photons. Overall, the development of quantum-light sources is now moving from conventional bulk crystals towards engineered quasi-phase-matched, chirped and aperiodically poled materials, as well as integrated nonlinear waveguides. This evolution is enabling broader bandwidths, improved wavelength tunability, access to the medium and far-infrared spectral regions, and increasingly sophisticated spectroscopic and imaging procedures.
For the spectroscopical applications, the non-degenerate sources represent the most interesting case since they allow us to generate pairs of correlated photons emitted at different wavelengths and link spectral ranges, e.g. the infrared with the visible one.
In this paper we will recall the nonlinear processes mainly employed to generate quantum states of light, we describe the state of the art of the sources used for several applications in the spectroscopical domain and propose a novel approach based on Chirped Apodized potassium titanyl phosphate (caKTP) crystals. The advanced engineering of these innovative crystals allows us to work with a source of quantum light characterized by a broad spectral width and a flat intensity distribution overcoming and significantly improving the drawbacks showed by the sources available so far. They are also characterized by a large-aperture and this distinctive feature represents a relevant novelty since it has been implemented also with short poling periods. Here we discuss theoretical simulations and the results obtained within an extensive experimental testbed based on the framework of the Quantum Ghost Spectroscopy (QGS) \cite{Chiuri2025}. Although this work focuses specifically on the near-infrared range, it also demonstrates the achievable performances paving the way to a wide range of possible applications, since the proposed technology could represent a valuable tool for the Quantum Technologies.

\section{Nonlinear Processes for Sources of Discrete Quantum States of Light}

The most widely employed platforms are based on the following nonlinear processes happening in several materials and highly engineered crystals \cite{Boyd}: Spontaneous Parametric Down-Conversion (SPDC), Spontaneous Four-Wave Mixing (SFWM), and, more recently, stimulated parametric processes. While for decades most of the efforts have been devoted to the study of the physical processes and their exploitation for quantum state engineering, nowadays the focus has been moved to the realization of advanced materials, structures and configurations for high-performances sources of quantum light \cite{Arahata21,Tashima24,Hashimoto25,Paterova2022,Hojo2021,Faleo26}. 

The SPDC still represents the main resource for both the degenerate and the non-degenerate regimes. When a laser beam impinges a bulk crystal, 2 photons are generated according to the Phase Matching (PM) conditions ensuring momentum and energy conservation:

\begin{equation}
    \omega_p = \omega_s + \omega_i , \ k_p = k_s + k_i 
\end{equation}
where $k_{p,s,i}$ and $\omega_{p,s,i}$ are the wavenumbers and frequency of the three beams, i.e. the pump (p), signal (s) and idler (i). 
These should be adapted for material involving the Quasi Phase Matching (QPM) conditions in modulated nonlinear media. In this case a term must be introduced, i.e. the mismatch $\Delta k$, that is linked to the modulation:

\begin{equation}
    \Delta k = k_p - k_s - k_i = \frac{2\pi}{\Lambda} 
\end{equation}

These sources rely on multilayer crystals that can be suitably engineered manipulating the size and shape of each single subdomain constituting the whole crystal. 

The third approach is based on the generation of photons by the SFWM in optical fibers or Photonic Integrated Circuits (PICs). The SFWM is a $\chi^{(3)}$ process where two photons from a pump laser are annihilated to create a signal photon of higher frequency and an idler photon of lower frequency. The generated frequencies are dictated by the usual conservation laws, which for co-polarized photons read as: 

\begin{equation}
    2 \hbar\omega_p = \hbar\omega_s + \hbar\omega_i , \ 2k_p = k_s + k_i + 2\gamma P_p
\end{equation}

Here the final term accounts for self- and cross-phase modulation caused by the strong pump pulse, which depends on the nonlinear coefficient of the fiber $\gamma$ and the peak power of the pump $P_p$.
There are 16 different phase-matching configurations available, depending on which of the four photons involved propagate along the fast or slow axes. This allows different phase-matched wavelengths to be accessed by selecting the desired pump and collection polarizations.

\section{Quantum States of Light for Spectroscopical Applications: State of the Art}

\textit{Quantum spectroscopy -} Here we include the QGS, Heralded Spectroscopy, Quantum Infrared Spectroscopy (QIRS), and Nonlinear Interferometry (NLI) that have been have built on SPDC sources. Early demonstrations employed bulk nonlinear crystals such as BBO, $MgO:LiNbO_3$, LBO, and $LiIO_3$, providing tunable photon pairs over spectral ranges of several tens of nanometers. The spectral bandwidth and central wavelength can be controlled through crystal thickness and phase-matching conditions \cite{Yabushita04, Chiuri22, Chiuri24, Kurita25, Scarcelli, Kalachev07, Kalachev_2008}.

Multi-crystal configurations based on BBO and $MgO:LiNbO_3$ have subsequently enabled broader and more flexible emission \cite{Arahata21,Kalashnikov14prx}, while periodically poled nonlinear crystals (ppNLs), e.g. ppLN, ppKTP and stoichiometric lithium tantalate (SLT), have provided efficient quasi-phase matching and convenient solutions \cite{Tashima24,Sansoni25,Sherwani26}. To further extend the accessible bandwidth and enable broadband non-degenerate photon-pair generation, including in the mid-infrared (MIR), sources of quantum light have been realized through temperature tuning, spatial displacement - fan-out crystals \cite{Hojo2021,Dolasinski15} - , and extensive engineering of innovative PPNLs - e.g. chirped and aperiodically poled structures \cite{Hashimoto25,Zotti25}.

To access longer wavelengths, materials such as $AgGaS_2$ (AGS) and $ZnGeP_2$ (ZGP) have been investigated because of their broad transparency windows and high nonlinear coefficients, enabling photon-pair generation in the fingerprint and far-infrared regions \cite{Paterova2022,Mukai:22, He26}. Alternative sources based on SFWM in birefringent fibers\cite{Pearce20}, photonic-crystal fibers \cite{Starshinov2024}, and silicon-on-insulator waveguides \cite{Sanna24} offer broadband and highly non-degenerate photon pairs, while intermodal phase matching can reduce Raman noise and provide efficient tuning of MIR photons.

\textit{Open-path Quantum Spectroscopy - } Open-path quantum spectroscopy combines quantum-light sources with free-space sensing and is particularly promising for atmospheric trace-gas detection. PPLN-based SPDC sources have been used to generate NIR–MIR photon pairs, with MIR wavelengths around 3–4 $\mu m$ suitable for molecular spectroscopy \cite{Neves26, Neves24}. Stimulated parametric down-conversion (ST-PDC) provides an alternative approach based on stimulated nonlinear interactions based on three-wave mixing of pump, signal, and idler modes after a double pass through a nonlinear crystal, producing narrowband MIR radiation and enabling nonlinear interferometric detection of gases such as methane \cite{Dong25aplphot, Dong25apl, Cardoso24}.

\textit{Quantum Hyperspectral and Multi-DOF Measurements - } Quantum hyperspectral imaging exploits the simultaneous spatial and spectral correlations of photon pairs to obtain information about both the position and spectral response of a target. Bulk BBO \cite{Chiuri23}, ppLN and ppKTP \cite{Paterova20, Zhang23, Placke26} sources have been used to demonstrate hyperspectral measurements from the visible to the MIR. More recently, aperiodically poled KTP (aKPT) has enabled hyperentangled states involving multiple degrees of freedom, such as frequency and polarization, opening the possibility of performing complementary measurements within a single quantum source \cite{Faleo26}.

\textit{Other Quantum-Inspired and Quantum-Enhanced Spectroscopies - } Quantum-enhanced and quantum-inspired approaches have also been explored for THz spectroscopy and Raman spectroscopy. ppLN-based SPDC can generate correlated visible–THz photon pairs \cite{Kutas21}, while optical rectification in materials such as GaP provides an alternative route to intense single-cycle THz radiation \cite{Adamou25}. In quantum-enhanced stimulated Raman scattering (QE-SRS), amplitude-squeezed light generated using a ppLN-based optical parametric amplifier is employed to reduce quantum noise and improve the sensitivity of Raman measurements \cite{Akatev26}.
Another promising approach is MIR heralded single-photon spectroscopy. In this scheme, MIR photons generated by SPDC interact with the sample and are subsequently upconverted to the NIR, where efficient Si-based single-photon detectors can be used. Chirped periodically poled crystals provide broadband MIR emission and allow spectrally resolved measurements through dispersive time-to-frequency mapping \cite{Cai24}.

\textit{Remote Measurements - } Quantum correlations can also be exploited for measurements performed over long distances. In one demonstration, frequency-correlated photon pairs generated by SFWM in a silicon nanowire were distributed between two locations, with one photon traveling through 50 km of optical fiber. The spectral information encoded in one photon could therefore be correlated with measurements performed remotely on its partner, demonstrating the potential of quantum-correlated light for distributed and remote sensing \cite{Dong2016}.

\begin{table}[!htbp] 
\centering 
\caption{Overview of quantum-light sources and nonlinear platforms employed for different spectroscopic and sensing applications.} \label{tab:quantum_sources_applications} 
\footnotesize 
\renewcommand{\arraystretch}{1.3} 
\makebox[\textwidth][c]{ 
\begin{adjustbox}{width=1.15\textwidth} 
\begin{tabular}{ >{\raggedright\arraybackslash}p{0.15\textwidth} >{\raggedright\arraybackslash}p{0.19\textwidth} >{\raggedright\arraybackslash}p{0.20\textwidth} >{\raggedright\arraybackslash}p{0.19\textwidth} >{\raggedright\arraybackslash}p{0.22\textwidth} } \hline \textbf{Application} & \textbf{Nonlinear Process} & \textbf{Source platforms} & \textbf{Spectral range} & \textbf{Main advantage} \\ \hline 

\textbf{Spectroscopy: QIRS, QGS, Heralded Spectroscopy, NLI} & SPDC, SFWM \cite{Yabushita04, Chiuri22,Chiuri24,Kurita25,Scarcelli,Kalachev07,Kalachev_2008,Kalashnikov14prx,Arahata21,Sansoni25, Tashima24,Sherwani26,Dolasinski15,Hojo2021,Zotti25,Hashimoto25,Mukai:22, Paterova2022,He26,Pearce20,Starshinov2024} & BBO, $MgO : LiNbO_3$, LBO, $LiIO_3$, PPLN, PPKTP, PPSLT, AGS, ZGP, birefringent and photonic-crystal fibers, SOI waveguides & Visible $\rightarrow$ MIR/FIR; broadband and tunable photon-pair generation & Spectral correlations over different spectral ranges and access to the fingerprint regions \\

\textbf{Open-path quantum spectroscopy: NLI} & SPDC, ST-PDC \cite{Neves24,Neves26,Dong25aplphot,Dong25apl,Cardoso24} & PPLN and MgO:PPLN & NIR--MIR & Remote/free-space trace-gas sensing \\ 

\textbf{Quantum hyperspectral, multi-DOF measurements} & SPDC  \cite{Chiuri23, Paterova20, Zhang23, Placke26, Faleo26}  & BBO, PPLN, PPKTP, aKTP & Visible $\rightarrow$ MIR & Joint measurements on multiple DOFs \\

\textbf{Quantum-enhanced / inspired spectroscopy: THz, QE-SRS, heralded MIR spectroscopy} & SPDC \cite{Kutas21,Adamou25,Akatev26,Cai24} & PPLN, GaP, PPLN waveguides, chirped PPLN & Visible $\rightarrow$ NIR/MIR; THz & Reduced quantum noise, enhanced sensitivity, and efficient MIR detection \\

\textbf{Remote measurements} & SFWM \cite{Dong2016} & Silicon nanowires and optical fibers & Telecom $\sim$1.5~$\mu$m & Long-distance distribution of quantum correlations for remote and distributed sensing \\ \hline 
\end{tabular} 
\end{adjustbox} } 
\end{table}


\section{Chirped Apodized Crystals} 
In a chirped QPM grating, the local poling period varies along the propagation direction. Consequently, different signal and idler wavelengths fulfil the QPM condition at different positions within the crystal. Chirping therefore broadens the phase-matching bandwidth compared with a uniform grating of the same length.

In the low-gain regime, the generated field can be described by the spatial Fourier transform of the nonlinear coefficient profile:

\begin{equation}
\Phi(\Delta k)\propto
\int_{0}^{L} d(z)\exp(i\Delta k z)\mathrm{d}z,
\end{equation}
where $d(z)$ represents the spatially modulated nonlinear coefficient and $\Delta k=k_p-k_s-k_i$ ($k_l,\ l=i,p,i$ indicates linear momentum of pump, signal and idler respectively) is the phase mismatch. 

In a finite crystal, the nonlinear polarization is abruptly switched on at the input facet and switched off at the output facet. As a consequence of the Fourier transform, these sharp spatial boundaries of the nonlinear coefficient cause ripples in the generated spectrum.
These ripples can be suppressed by apodizing the QPM structure, thereby replacing the abrupt boundaries of the effective interaction with gradual transitions. In the present work, this is achieved using nonlinear-chirp apodization. Within the central part of the crystal, the grating vector varies linearly along the crystal length, providing the desired phase-matching bandwidth. Near the input and output facets, the grating vector follows closely the nonlinear-chirp apodization profile introduced in \cite{Phillips2013}. In these regions, the chirp rate increases smoothly and monotonically towards the crystal boundaries, moving the grating vector rapidly away from the phase-matching range.

\begin{figure}[H]
\centering

\begin{minipage}[t]{0.47\linewidth}
    \centering
    \includegraphics[width=\linewidth]{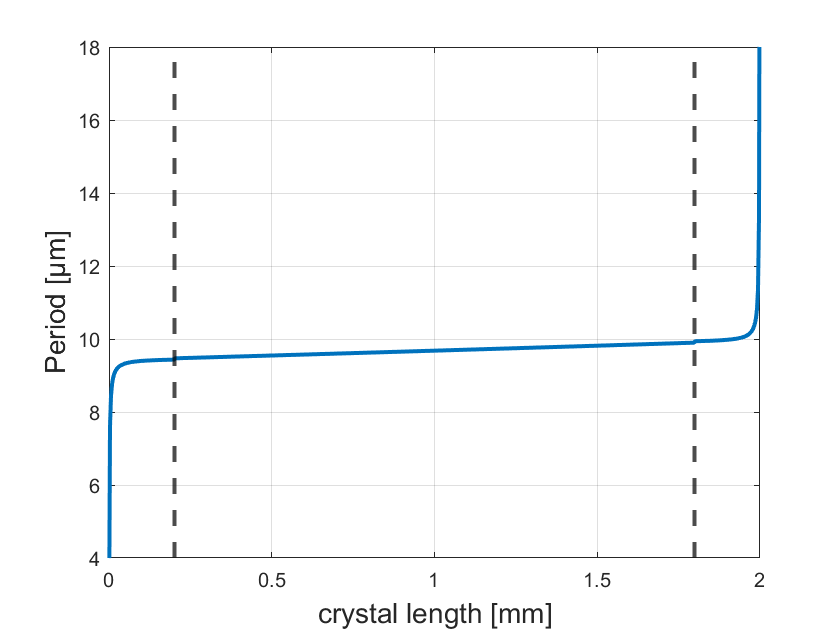}\\
    \textbf{(a)}
\end{minipage}
\hfill
\begin{minipage}[t]{0.47\linewidth}
    \centering
    \includegraphics[width=\linewidth]{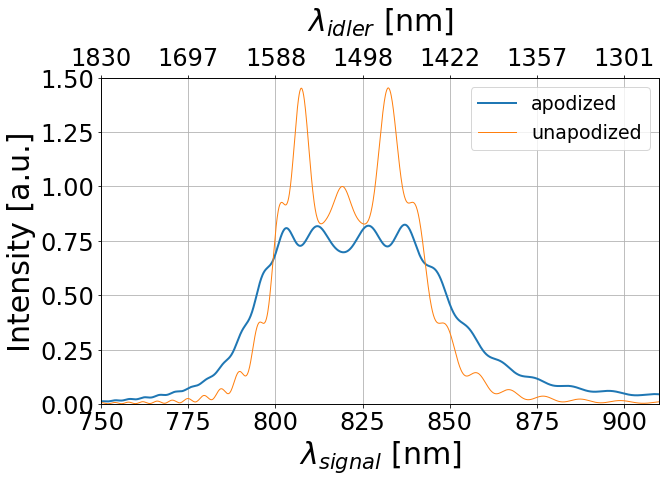}\\
    \textbf{(b)}
\end{minipage}

\medskip

\begin{minipage}[t]{0.47\linewidth}
    \centering
    \includegraphics[width=\linewidth]{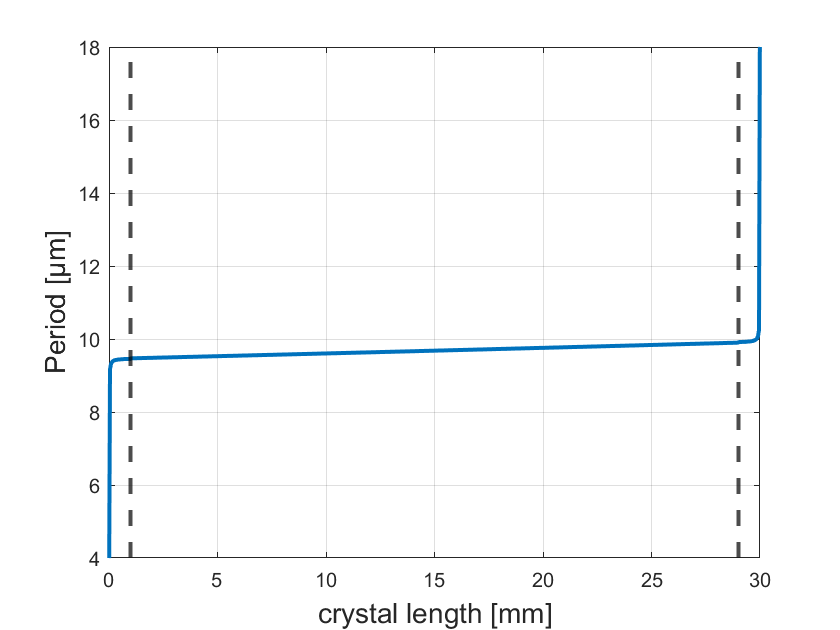}\\
    \textbf{(c)}
\end{minipage}
\hfill
\begin{minipage}[t]{0.47\linewidth}
    \centering
    \includegraphics[width=\linewidth]{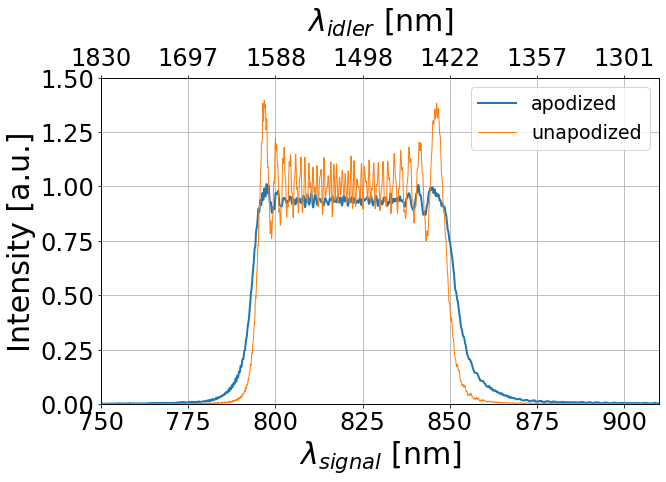}\\
    \textbf{(d)}
\end{minipage}

\caption{Effect of nonlinear-chirp apodization on the QPM structures and their calculated phase-matching spectra. (a) Poling-period profile for the 2~mm crystal. The black dotted line indicates the transition from linear chirp to apodization. (b) Corresponding phase-matching spectra. (c) Poling-period profile for the 30~mm crystal. The black dotted line indicates the transition from linear chirp to apodization. (d) Corresponding phase-matching spectra.}
\label{fig:chirp_apodization}
\end{figure}

This creates a gradual transition between the phase-matched central region and the strongly phase-mismatched regions near the facets. The spectral response is therefore smoothly tapered rather than abruptly truncated, suppressing interference ripples and producing a flatter spectrum. All simulations presented below assume type-0 SPDC in chirped apodized KTP (caKTP) at {room temperature}, 
with the pump, signal, and idler fields vertically polarized. This configuration accesses the largest effective nonlinear coefficient of KTP and therefore provides the highest conversion efficiency. Figure~\ref{fig:chirp_apodization} illustrates the effect of nonlinear-chirp apodization for two crystal lengths: a 2~mm design corresponding to the crystal used in the experiment and a 30~mm design representing a longer interaction length. Both designs have a nominal poling-period range from
$9.49~\mu\mathrm{m}$ to $9.90~\mu\mathrm{m}$.
For a pump wavelength of $532~\mathrm{nm}$, these periods correspond
to signal--idler wavelength pairs of approximately $850~\mathrm{nm}$
and $1422~\mathrm{nm}$ at $9.49~\mu\mathrm{m}$, and $793~\mathrm{nm}$
and $1616~\mathrm{nm}$ at $9.90~\mu\mathrm{m}$.

Figure~\ref{fig:chirp_apodization}(a) shows the  apodized poling-period profile for the 2~mm design. The grating follows the nominal linear chirp between 0.2 and 1.8~mm along the crystal, while nonlinear-chirp apodization is applied over the first and final 0.2~mm of the crystal. This apodized profile corresponds to the crystal investigated experimentally in Section~\ref{sec:experimental}. The simulated spectra are shown in Figure~\ref{fig:chirp_apodization}(b). The red curve shows the simulated spectrum produced by the unapodized grating, which is linearly chirped over the entire 2~mm crystal length, whereas the blue curve shows the spectrum produced by the nonlinear-chirp-apodized grating. Both curves are normalized by same normalization factor. The relative reduction in the peak spectral intensity therefore represents the small efficiency penalty associated with apodization, caused by a smaller interaction length. 

Figures~\ref{fig:chirp_apodization}(c) and (d) show the corresponding poling-period profile and simulated spectra for the 30~mm design. In the apodized case, the grating follows the nominal linear chirp between 1 and 29~mm along the crystal, while nonlinear-chirp apodization is applied over the first and final ($1~mm$). The unapodized grating is linearly chirped over the entire $30~mm$ crystal length. The 30 mm long crystals have a brightness as total pair rate that is approximately 15 times higher compared to the 2 mm crystals. For both crystal lengths, apodization strongly suppresses the spectral oscillations while preserving approximately the same useful phase-matching bandwidth. For comparison, uniform caKTP gratings designed for a signal wavelength of $810~nm$ have calculated bandwidths of $5.9~nm$ and $0.4~nm$ for crystal lengths of $2~mm$ and $30~mm$, respectively. These values illustrate the substantially broader spectral coverage of the chirped designs, which extends from approximately $793$ to $850~nm$.

To investigate the applicability of the proposed design approach beyond the experimentally studied visible–near-infrared wavelength range, an additional caKTP grating was designed for strongly nondegenerate SPDC with one photon in the visible and the other in the mid-infrared. The same design and apodization procedure described above were applied and is therefore not repeated here.

\begin{figure}[H]
    \centering
    \includegraphics[width=0.8\linewidth]{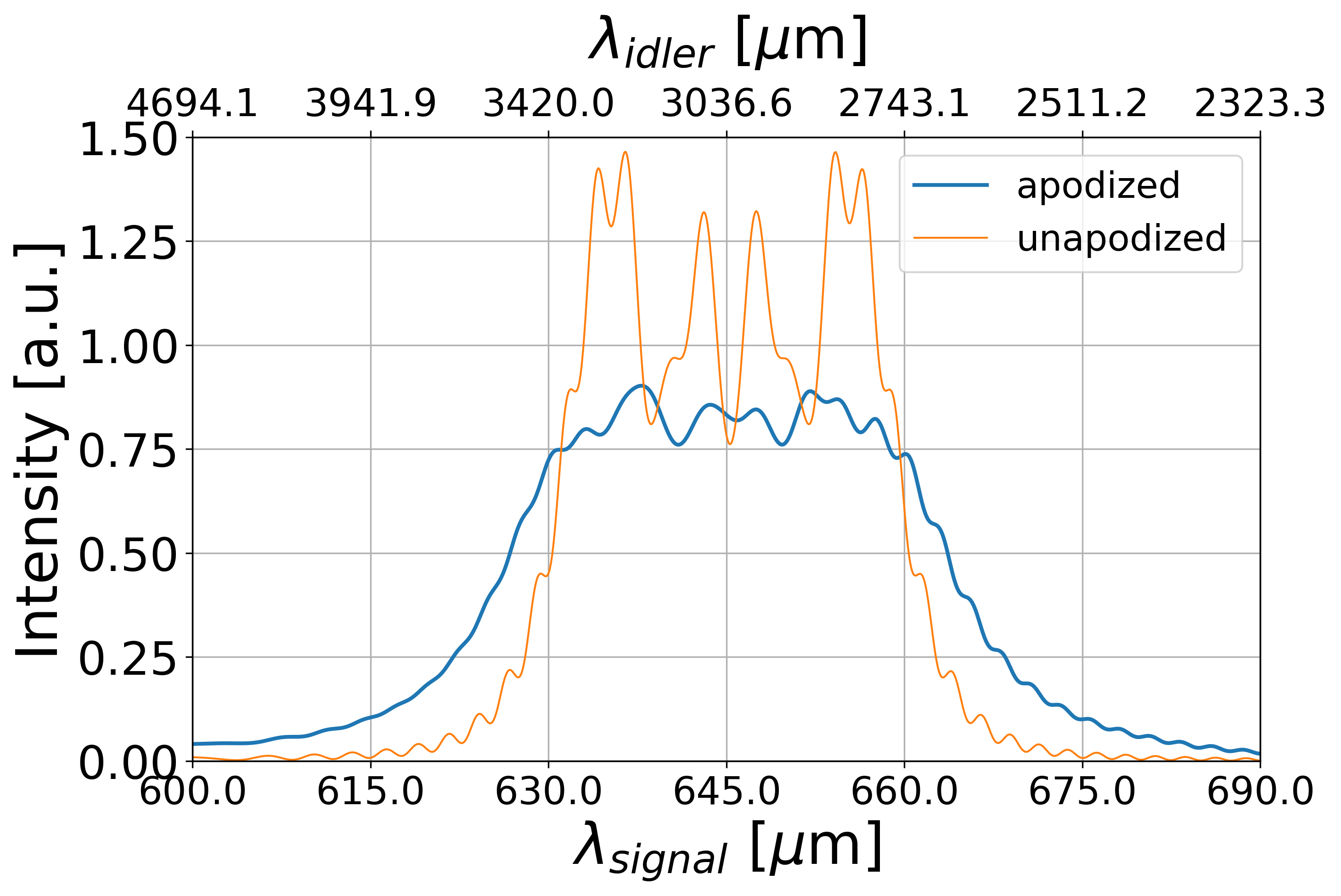}
    \caption{Simulated spectrum of a 2~mm-long nonlinear-chirp-apodized ppKTP crystal pumped at 532~nm. The nominal poling period varies from 12.38~$\mu$m to 13.71~$\mu$m, and the resulting idler spectrum extends from approximately 2.6~$\mu$m to 3.6~$\mu$m.}
    \label{fig:mir_simulation}
\end{figure}

Figure~\ref{fig:mir_simulation} shows the predicted output spectrum for a 2~mm-long crystal pumped at 532~nm. The grating is nominally chirped from a poling period of
\SI{12.38}{\micro\metre} to \SI{13.71}{\micro\metre}.
A period of \SI{12.38}{\micro\metre} corresponds to idler and signal
wavelengths of approximately \SI{2.7}{\micro\metre} and
\SI{663}{\nano\metre}, respectively, while a period of
\SI{13.71}{\micro\metre} corresponds to idler and signal wavelengths
of approximately \SI{3.5}{\micro\metre} and \SI{627}{\nano\metre},
respectively.
A relatively short crystal was selected to limit the influence of propagation losses in a future experimental implementation. The simulation demonstrates the potential of nonlinear-chirp-apodized ppKTP for generating broadband visible–MIR photon pairs with a smooth spectral response. This design represents a possible extension of the experimentally investigated source and was not fabricated or characterized in the present work.
\section{Experimental Characterization}
\label{sec:experimental}
\begin{figure}[H]
    \centering
    \includegraphics[width=0.75\linewidth]{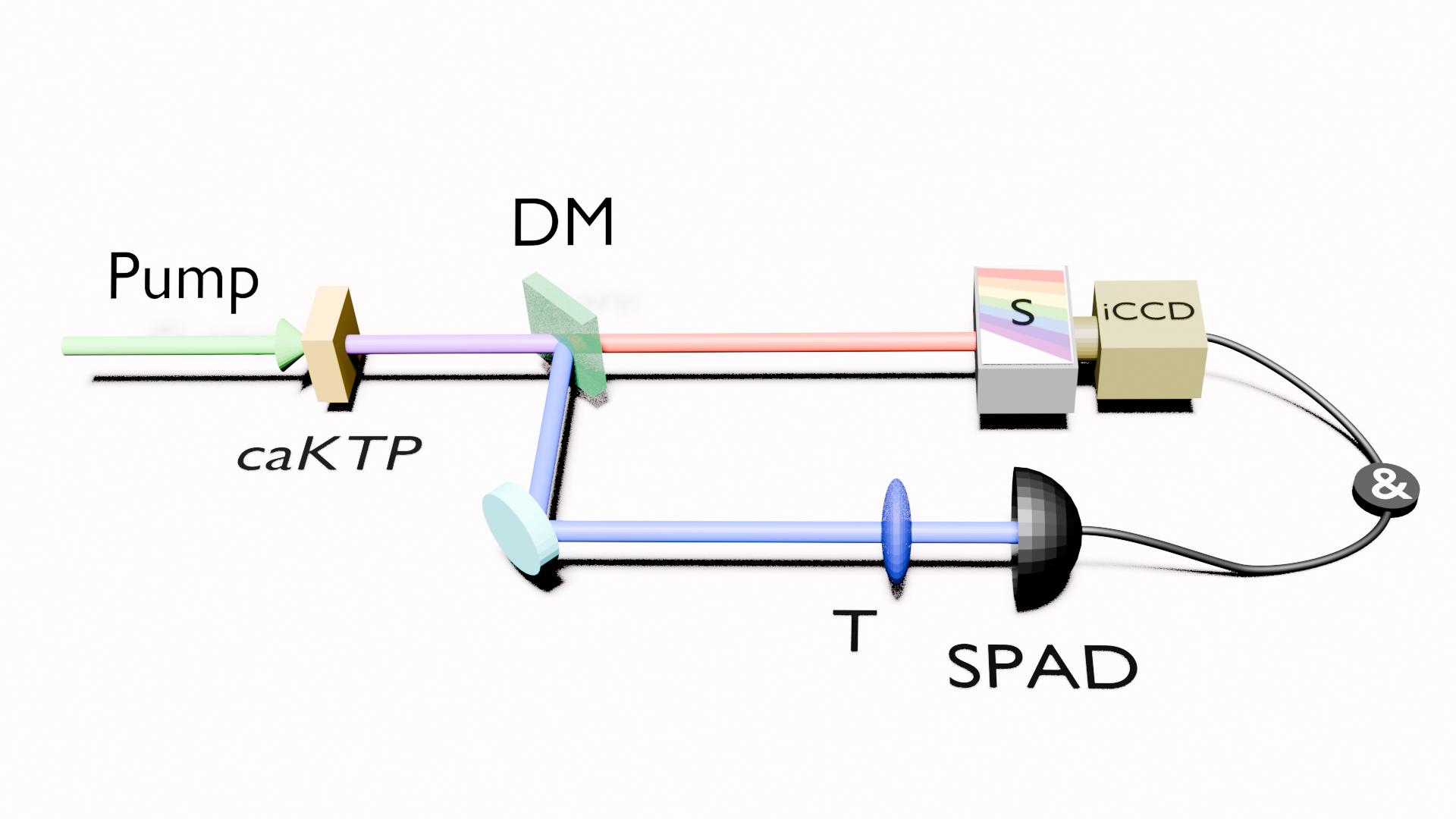}
    \caption{Experimental Setup for characterization of the chirped apodized source. Here a pulsed laser at 532 nm (8 ps pulses with PRR=40 MHz) impinges on the caKTP crystal. Pairs of single photons in the VIS and IR range are produced and split on a dichroic mirror (DM). The IR photon goes through a target sample (T) and is then detected by a single pixel SPAD; the VIS photon is spectrally analyzed by means of a spectrograph (S) and an intensified CCD camera (iCCD). Coincidences between the detector and the camera are recorded (\&).}
    \label{fig:setup}
\end{figure}

The characterization of the source has been carried out in the setup shown in Fig. \ref{fig:setup} and in the low-gain regime. Here a pulsed laser at 532 nm with pulses of 8 ps, 40 MHz repetition rate (PRR) and a variable average power - $\approx 0-250 mW$ - impinges on a caKTP crystal (either 1 or 2mm long) characterized by a high aperture $4 \times 4$ mm at temperature of $40^{\circ} C$. SPDC pairs are generated in a highly nondegenerate regime. The two photons, one in the telecom range and the other one centered around 810 nm, are then split on a dichroic mirror (DM). Before being measured with a single pixel SPAD detector, the telecom photon may travel through a target, either an interference filter or a long pass filter. The visible photon is instead spectrally analized by means of a spectrometer (Andor Kymera 328I-A-SIL) and detected on an intensified CCD camera (Andor iSTAR iCCD DH334T-18U-73). To record correlated events a click on the IR detector triggers the acquisition window of the camera, such that it only can acquires signal when a photon is detected on the Telecom detector.

\subsection{Bandwidth and brightness}
\begin{figure}[H]
\centering
\begin{minipage}[t]{0.47\linewidth}
    \centering
    \includegraphics[width=\linewidth]{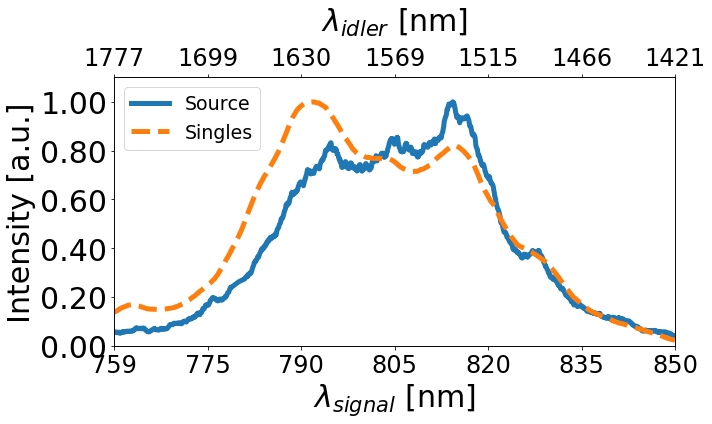}\\
    \textbf{(a)}
\end{minipage}
\hfill
\begin{minipage}[t]{0.47\linewidth}
    \centering
    \includegraphics[width=\linewidth]{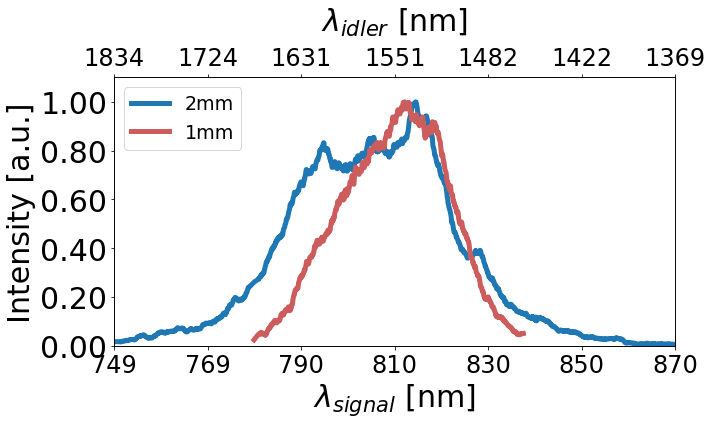}\\
    \textbf{(b)}
\end{minipage}
\caption{(a) Spectrum of a 2mm-long caKTP. The blue curve reports the correlated spectrum, while the orange dashed line refers to the uncorrelated measurements where the background - counts when the SPDC source is blocked - has been subtracted. The reduced spectral width of the correlated spectrum is due to limited detection efficiency of the telecom detector for wavelengths above 1600 nm. 
(b) Correlated spectra of 1mm- (red curve) and 2mm-long crystals (blue curve). It is clear that the longer the crystal, the broader the spectrum. In these particular cases we have a FWHM of $\approx 28nm$ (signal) / $\approx 103 nm$ (idler) for the 1mm crystal, while the 2mm source produces bandwidths of $\approx 36nm$ (signal) and $\approx 138 nm$ (idler).
}
\label{fig:source}
\end{figure}

We first characterize the spectrum of the SPDC light generated in a 2-mm-long crystal. The measurements are performed with the setup shown in Fig. \ref{fig:setup} where the target sample is removed from the idler path. The collected data are reported for both correlated (blue line) and uncorrelated (orange dotted line) detection events in Fig. \ref{fig:source} (a). The correlated spectrum is obtained by recording only coincidence events between the iCCD camera and the SPAD, whereas the uncorrelated spectrum is measured by collecting all photon counts detected by the intensified camera in a free-running configuration.

The measurements clearly show the broad spectral bandwidth of the SPDC emission in agreement with the simulations reported in Fig. \ref{fig:chirp_apodization} (b). The correlated spectrum appears narrower than the uncorrelated spectrum, primarily due to the reduced detection efficiency of the SPAD at wavelengths beyond 1600 nm \cite{mpd} and to possible spectral filtering that can happen inside the crystal, which is caused by some blocking mechanism and which could be ascribed to scattering on the inhomogeneities appearing in the caKPT crystal because of fabrication imperfections [See the Supplementary Material of \cite{Hashimoto25}]. This decrease in efficiency leads to a lower number of detected idler photons and, consequently, to a reduced number of triggers available for the iCCD camera, limiting the acquisition of correlated events in this spectral region.
It is well known that the spectral bandwidth depends on the length of the nonlinear crystal. We therefore characterized the spectrum of Fig. \ref{fig:source} (a) with the one generated by a shorter, 1-mm-long ppKTP crystal. Although the spectral bandwidth usually decreases with increasing length, for caKTP crystals, such as the ones used in our experiment, the opposite trend is observed: the spectral bandwidth increases with the crystal length, as shown in the simulation of Fig. \ref{fig:chirp_apodization}. This behavior experimentally obtained as evident in Fig. \ref{fig:source} (b), which shows the correlated spectra measured using the 1-mm-long (red line) and 2-mm-long (blue line) crystals.

For the 1-mm-long crystal, the measured FWHM is approximately 28 nm in the visible range and 103 nm in the telecom band. In comparison, the 2-mm-long crystal exhibits broader spectra, with FWHM values of approximately 36 nm and 138 nm in the visible (signal) and Telecom (idler) ranges, respectively.
Further crystals with higher lengths have been not tested because of the limited pulse duration of the laser. 

Such a broad bandwidth is accompanied by high brightness, a feature that is relevant both for ghost spectroscopy experiments, in which photon pairs are exploited, and for nonlinear interferometry setups, where single photons are instead the key resource. We characterized the brightness of our sources in terms of both pair and single-photon generation rates. To this end, we measured the coincidence rates and normalized them to the Klyshko efficiency \cite{Klyshko1980} of the setup and to the pump power. We obtained correlated-pair brightness values of $\mathcal{B}^{corr}_{2\mathrm{mm}} = 4.8\cdot 10^4$ pairs/(s·mW) for the 2-mm-long crystal and $\mathcal{B}^{corr}_{1\mathrm{mm}} = 2.5 \cdot 10^4$ pairs/(s·mW) for the 1-mm-long source.
For the single-photon brightness in the visible range, we obtained $\mathcal{B}^{sing}_{2\mathrm{mm}} = 3.1 \cdot 10^5$ photons/(s·mW) and $\mathcal{B}^{sing}_{1\mathrm{mm}} = 2.1 \cdot 10^5$ photons/(s·mW), respectively [See Section \ref{discussion} for a complete comparison with the existing sources]. The overall detection efficiencies of the setup were estimated to be $\eta_{VIS}\approx 0.3$ and $\eta_{NIR}=0.08$, mainly limited by coupling - that for highly nondegenerate photons becomes crucial - and detector efficiencies. In particular, the latter is strongly affected by the low detection efficiency of the NIR detector, which is approximately 15\%.

\begin{figure}[H]
\centering
  \includegraphics[width=0.75\linewidth]{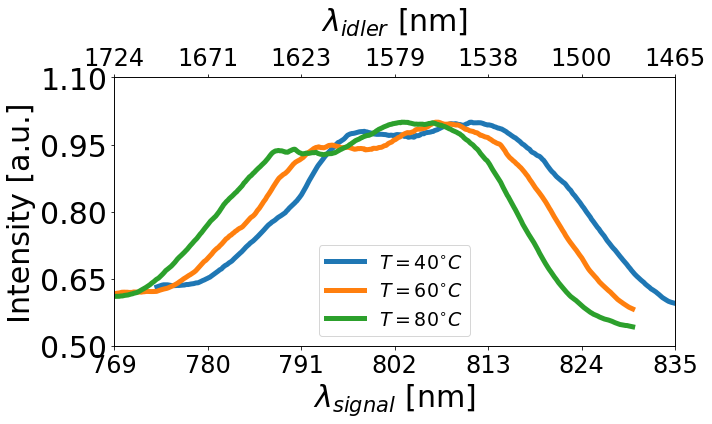}
  \label{fig:Singles1mm}
\caption{Uncorrelated spectrum of a 1mm-long caKTP for different temperatures, namely $40^{\circ} C$ (blue curve), $60^{\circ} C$ (orange curve) and $80^{\circ} C$ (green curve). Increasing the temperature the signal spectrum shifts toward lower wavelengths, while the idler spectrum toward higher wavelengths.
}
\label{fig:temperature}
\end{figure}

Our source also offers a degree of tunability of the central wavelength of the emitted spectrum. In particular, the central emission wavelength can be tuned by varying the crystal temperature. Figure \ref{fig:temperature} shows the emission spectra of a 1-mm-long caKTP crystal measured at different temperatures. As the temperature increases, the spectrum shifts towards higher wavelengths for the signal and towards lower wavelengths for the idler.

\subsection{Coincidence-to-Accidental Ratio (CAR)}
A meaningful figure of merit for characterizing a photon-pair source is the coincidence-to-accidental ratio (CAR) \cite{Sansoni25}. The CAR quantifies the contribution of correlated photon pairs relative to the unavoidable contribution from uncorrelated detection events. A higher CAR therefore indicates a higher-quality source. The uncorrelated contribution mainly arises from multi-pair generation events. This figure of merit is particularly relevant in experiments involving correlated photon pairs, where a sufficiently high CAR is required. Typically, a $CAR > 6$ is considered a useful threshold for pair sources in quantum technology applications \cite{Pitsch:21,Pitsch23,Pitsch25,Sansoni25}.

\begin{figure}[H]
\centering
\includegraphics[width=0.75\linewidth]{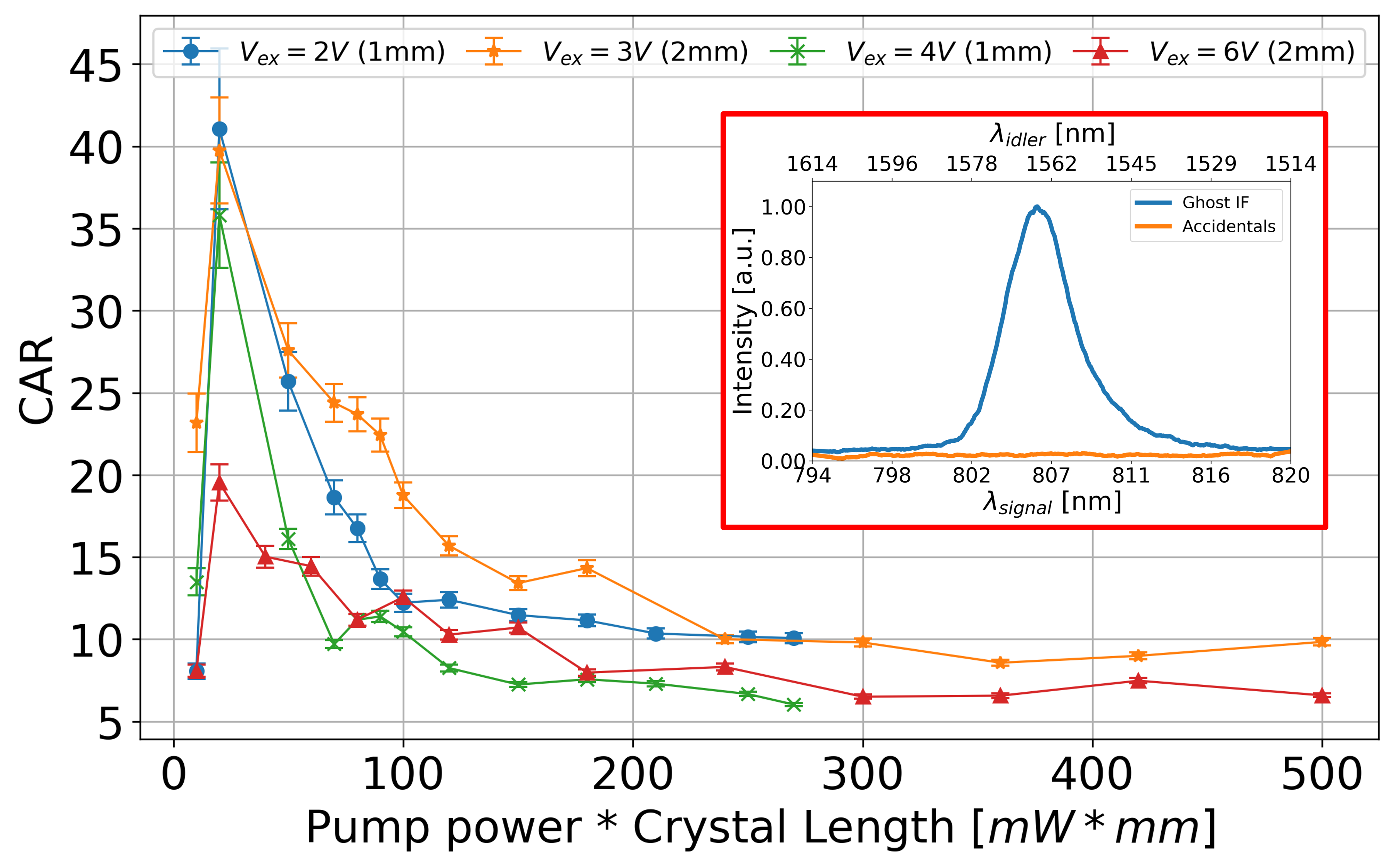}
\caption{CAR of a QGS measurement as a function of pump power per crystal length for different bias voltages $V_{ex}$ of the SPAD detector. Inset: quantum ghost spectroscopical measurement of an interference filter centered at 1550 nm collected with a pump power of 80mW on a 1mm-long crystal corresponding to a CAR of 9.85. Shift in the central wavelength of the ghost spectrum is due to a shift in the camera calibration, for details refer to \cite{Chiuri24}.
}
\label{fig:CAR}
\end{figure}

We therefore characterized the CAR of our sources. To this end, we performed a quantum ghost spectroscopy (QGS) experiment \cite{Chiuri24} using the setup shown in Fig. \ref{fig:setup}, with an interference filter (central wavelength 1550 nm FWHM 10 nm) placed as the target in the Telecom arm. We varied the pump power and collected the corresponding ghost measurements using the iCCD camera; an example of the retrieved data is shown in the inset of Fig. \ref{fig:CAR}. Measurements were performed at different pump powers and for different bias voltages of the SPAD detector \cite{Chiuri21sens}, ranging from 2 V to 6 V, using both 1-mm- and 2-mm-long caKTP crystals. From such measurements we calculated the CAR and report it in Fig. \ref{fig:CAR}: here we observe the expected behavior of a typical CAR curve. At low pump powers, the CAR is relatively low because detector dark counts dominate the measured signal. As the pump power increases and the contribution from photon-pair generation becomes larger than the dark-count contribution, the CAR increases and reaches a maximum. At higher pump powers, the CAR decreases again as the probability of multi-pair generation events increases, leading to a larger contribution from accidental coincidences.

Figure \ref{fig:CAR} clearly shows this expected behavior. Notably, here the CAR remains above 6 over the entire investigated pump-power range, demonstrating the high quality of our source. Moreover, varying the SPAD bias voltage changes the height of the CAR peak. As expected, increasing the bias voltage increases the detector dark-count rate, thereby affecting the low-pump-power regime and the maximum CAR value. A good trade-off between dark counts and signal count rate can be obtained at a bias voltage of 3 V (orange markers and line), resulting in a high CAR across the investigated pump-power range.

This operating point allows us to use relatively high pump powers without significantly compromising the quality of the quantum correlations, while still maintaining a high count rate.

\subsection{Ghost spectroscopy of a LPF}
Following the characterization of our source, we exploit its broad spectral width and high CAR to demonstrate its potential for quantum ghost spectroscopy (QGS). We perform a QGS experiment using a long-pass filter (LPF) as the target, positioned at different angles with respect to the beam propagation direction. By varying the tilting angle, we modify the effective cut-off wavelength of the filter and, consequently, retrieve the corresponding transmitted spectra.

\begin{figure}[H]
\subfigure[]{
  \includegraphics[width=0.5\linewidth]{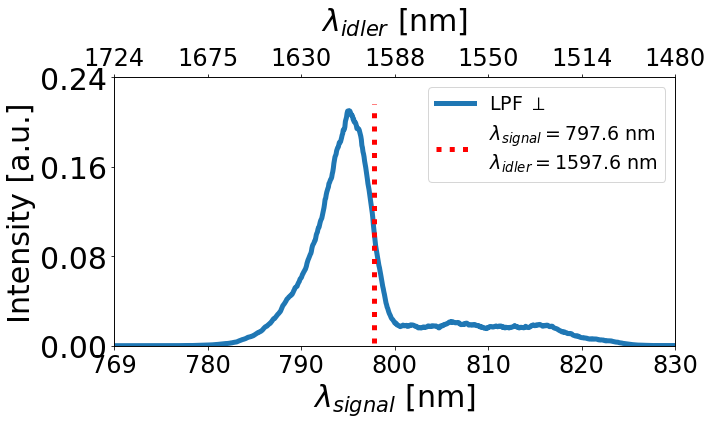}
  \label{fig:LPFperp}
}
\subfigure[]{
  \centering
  \includegraphics[width=0.5\linewidth]{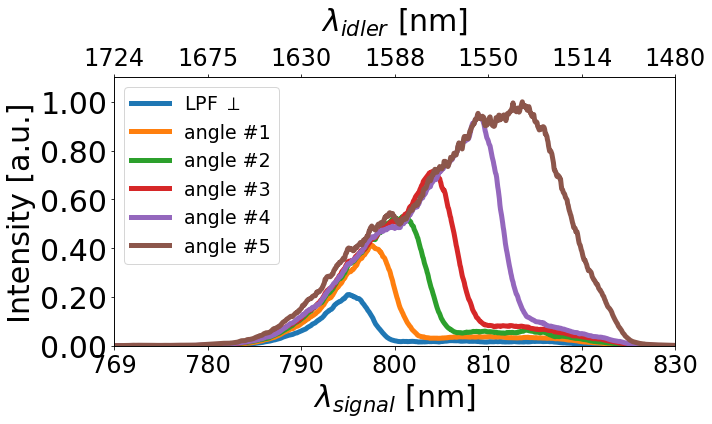}
  \label{fig:LPFtilted}
}
\caption{Ghost spectra of a long-pass filter (LPF) at different angles. (a) Ghost spectrum obtained when the LPF is perpendicular to the light beam. The observed cut-off wavelength agrees with the expected value. (b) Ghost spectra measured for different tilt angles of the LPF. As the tilt angle increases, the cut-off wavelength shifts toward longer (shorter) wavelengths for the signal (idler) photon. At sufficiently large tilt angles, the cut-off wavelength of the LPF moves beyond the spectral range of the source. In this regime, the QGS measurement reproduces the full correlated spectrum of the source.
}
\label{fig:LPF}
\end{figure}

We acquire spectra for six different LPF positions, with the results shown in Fig. \ref{fig:LPF}. When the LPF is positioned perpendicular to the beam, the retrieved spectrum is reported in Fig. \ref{fig:LPFperp}. The nominal cut-off wavelength of the LPF is 1583 nm, corresponding to a signal wavelength of 801 nm. Taking into account the calibration shift of the iCCD, $\delta\lambda=-4$ nm in the visible range, we expect the cut-off to occur at $\lambda_s=797$ nm, corresponding to an idler wavelength of $\lambda_i=1600$ nm. This is consistent with the result obtained experimentally (Fig. \ref{fig:LPFtilted}), with a CAR above 15 at a pump power of 100 mW. These conditions enable rapid measurements, with acquisition times on the order of tens of seconds per spectrum. 

By tilting the LPF, we retrieve different ghost spectra, as shown in Fig. \ref{fig:LPFtilted}. The results demonstrate that the spectral properties of our source are well suited for the QGS protocol, enabling both high-visibility measurements, as indicated by the large CAR, and the rapid acquisition of spectra with good spectral resolution.

\section{Discussion}
\label{discussion}
Both simulations and experimental measurements demonstrate that engineering a chirped and apodized KTP crystal provides a set of properties that make this class of quantum sources particularly well suited for quantum spectroscopy experiments. In Table \ref{tab:quantum_sources} we report the most promising sources proposed in literature and employed in the quantum spectroscopy domain demonstrating that the approach proposed here could be extremely fruitful also because our results have been obtained without focusing the pump. In addition, the spectral shape of a caKTP is not characterized by the ripples and fluctuations shown so far \cite{Tashima24,Vanselow19,Kaufmann22}.

\begin{table}[!htbp] 
\centering 
\caption{Overview of the most relevant existing sources of quantum light employed in the quantum spectroscopy domain.}
\label{tab:quantum_sources} 
\footnotesize 
\renewcommand{\arraystretch}{1.3} 
\makebox[\textwidth][c]{ 
\begin{tabular}{ >{\raggedright\arraybackslash}p{0.25\textwidth} >{\raggedright\arraybackslash}p{0.20\textwidth} >{\raggedright\arraybackslash}p{0.25\textwidth} >{\raggedright\arraybackslash}p{0.20\textwidth}  } \hline \textbf{Pump} & \textbf{Crystal and Regime} & \textbf{Generated Photons and Spectral Range} & \textbf{Notes} \\ \hline 

15 ps pulsed at 532-nm, PRR = 1 kHz, average power $\approx 1 mW$ (pulse energy $\approx 1\mu J$) \cite{Hashimoto25} & aperiodically poled LN (5 mm); high-gain regime  & Signal power 9 $\mu W$; 780-840nm / 1450-1650nm & focused on the crystal\\

15 ps pulsed at 532-nm, PRR = 1 kHz, average power $\approx 1 mW$ (pulse energy $\approx 1\mu J$) \cite{Zotti25} & aperiodically poled LN (5 mm); high-gain regime  & Signal power 9 nW; 2 crystals: centered at 3.3 and 2.6 $\mu m$ / 635 and 670 nm. Range for the first crystal: 620 - 650 nm / 3 – 3.8 $\mu m$ & focused on the crystal; 90 pW of MIR power probing the sample\\

720-nm CW, 400mW \cite{Sherwani26} & ppKTP (30 mm); low-gain regime  & $7.81 \times 10^8$ photons/second; 901–923 nm / 3.26–3.58 $\mu m$ & focused on the crystal\\

532-nm CW, 150mW \cite{Tashima24} & PPMg:SLT (10 mm); low-gain regime  & Signal photons: $8.0 \times 10^5$ counts/2s or $2.5 \times 10^4$ photons/mW; 590-725 nm / 2-5$\mu m$ & focused on the crystal, ripples in the source\\

785-nm CW, 100 mW \cite{Mukai:22} & AGS (0.5 mm); low-gain regime  & Signal photons: $< 1 \times 10^6$ counts/s (pairs estimated: $5 \times 10^7$ pairs/s - total detection efficiency: $2 \%$ -)  ; 840-920 nm / 5.4-12 $\mu m$ & focused on the crystal; to cover the whole range the crystal is tilted. Each curve has a bandwidth $ <1\mu m$ in the MIR range\\

638-nm CW, 50 mW \cite{Hojo2021} & 11 mm PPSLT fanout crystal; low-gain regime  & flux estimated by summing the photons detected from the whole structure: $1.4 \times 10^{10}$ photon/s; 700-800 nm / 2-5 $\mu m$ & focused on the crystal; fanout configuration\\
\hline 
\end{tabular} } 
\end{table}

Hence, the engineered phase-matching enables a very broad spectral emission, extending significantly into the infrared region and the brightness can be significantly improved by focusing the pump laser - collimated for our measurements - or increasing the length of the caKPT since the 30mm crystal shows a 15 times greater pair generation rate than the 2mm. 

On the other hand, the obtained results demonstrates that, through appropriate engineering of the crystal's chirp and apodization, the brightness is satisfactory also with short crystals. This enables the additional practical advantage of compatibility with pulsed pump lasers - rather than a continuous-wave (CW) pump - and pave the way to the use in sources of quantum light operating in the high-gain regime. 

Although the discussed specifications are highly desirable for spectroscopy applications, they could also impose stringent requirements on the detection scheme. In particular, the dark-count rate of the bucket detector can become a limiting factor for the efficiency of the measurement. Increasing the pump power can enhance the photon-pair signal above the dark-count background; however, at sufficiently high pump powers, the increased probability of multi-pair generation in SPDC introduces uncorrelated pairs, or colored noise, that progressively degrades the quantum correlations \cite{Sansoni25}. Consequently, an optimal operating regime must balance the need for a sufficiently high count rate against the preservation of strong correlations. The CAR measurements reported in Fig. \ref{fig:CAR} demonstrate that our source maintains a high level of correlations over a broad range of operating conditions, including configurations characterized by relatively high dark-count rates. This robustness against noise makes the caKTP source particularly attractive for quantum spectroscopy experiments in which the preservation of photon-pair correlations is essential. 


\section{Conclusion}
Our results demonstrate that engineering the chirp and apodization of KTP crystals enables the realization of quantum sources combining high brightness with exceptionally broad spectral emission into the infrared region.
These properties are complemented by a high CAR over a broad range of pump powers, demonstrating the robustness of the generated quantum correlations against detector dark counts and multi-pair noise. The combination of high brightness and the possibility of operating at relatively high pump powers enables efficient photon-pair collection and, consequently, faster spectroscopic measurements. In addition, the high aperture of these devices, makes them suitable candidates also for imaging and hyperspectral applications. Altogether, these characteristics make chirped and apodized KTP crystals a promising platform for fast quantum spectroscopy, not only limited to the QGS, as the schemes based on nonlinear interference, where broad bandwidth, high brightness, and strong quantum correlations are essential.

%
%

\ack{The Authors thank Marcello Nuvoli for the fabrication of mechanical holders.}

\funding{This work was supported by the NATO Science for Peace and Security
(SPS) Programme, project QUASAR (id G7883).}

\roles{Conceptualization, Data curation, Formal Analysis, Investigation, Methodology, Writing – review $\&$ editing, Review, and Editing: All Authors; Funding acquisition and Project administration: A.C.; Resources: M.W and P.M.; Supervision: A.C., Visualization:A.C., P.M., L.S., Writing – original draft: L.S. and P.M.}

\data{The data that support the findings of this study are available upon reasonable request from the authors.}


\bibliographystyle{iopart-num}
\bibliography{iopart-num}

\end{document}